# From Information to Affirmation: An Investigation on the Echo Chamber Effect from YouTube Comments under Technology Product Reviews

Hongrui Jin

kevin.jin@student.uva.nl

## Abstract

Social media may create echo chambers that reaffirm users' beliefs and opinions through repeated exposure of similar notions. Whilst the formation and effect of echo chambers have been intensively examined in thread-based platforms such as Twitter, Facebook and Reddit, we shift our focus on product review discussions on YouTube. This paper examines YouTube comments (n=2500) through a combined approach of quantitative content analysis (QCA) and sentiment analysis (SA) under selected selected YouTube videos (n=10). We conclude this paper by highlighting the formation of echo chamber effect in relation to comment argumentation and sentiments.

**Introduction**

*In Each Other We Trust: Echo Chambers*

Social media platforms such as YouTube, Twitter and Reddit have enabled users to share knowledge and opinions across time and locations. Billions of active users engage with both media contents and fellow users daily, exchanging direct messages, comments and other forms of information afforded by platforms. During this exchange of information on social media, users exhibit tendencies to favour narratives that aligns with their own beliefs (Del Vicario et al. 2016). This phenomenon is commonly referred to as the *echo chamber effect* . The *echo chamber* is further defined as an environment where opinions or beliefs of users are underpinned through repeated exposure of information akin to theirs (Cinelli et al. 2021). Since this mechanism poses influence exclusively on groups of individuals, scholars have warned the formation of echo chambers could affect how users consume media, radicalise and distort opinions, or worse, fragment societies by and large (Cinelli et al. 2021; Röchert et al. 2020). Whilst the formation and effect of echo chambers have been intensively examined in thread-based platforms such as Twitter, Facebook and Reddit (Van Raemdonck 2019; Quattrociocchi, Scala, and Sunstein 2016; Yusuf, Al-Banawi, and Al-Imam 2014), fewer studies focused on the presence of echo chambers on the audio-visual media platform YouTube. Additionally, previous studies on echo chambers predominately favours politics and public health utterances, while such effect could well exist in other discussions and shape user activities (Röchert et al. 2020).

*Like, Comment and Subscribe: YouTube as Social Media*

YouTube might not immediately resemble typical social media platforms such as Twitter and Instagram, yet at a closer look the platform establishes its social functionality by allowing user interactions such as content uploading, following/subscribing, (video) likes, and commenting. The social nature of video sharing platforms has been increasingly acknowledged by scholars, whereas some studies already noticed the potential of video comments breeding dominant opinions and views on politics, conspiracies and sexuality (Lutz et al. 2021; Naím 2007). In relation to echo chambers, we have identified two non-exclusive types of YouTube users: *content creators,* who initially present an environment (video upload) **and** topic for discussion, as well as *viewers* that audit and respond to information provided by *content creators* and other *viewers* in an established environment according to their personal experiences. While any user could assume both positions on the platform, any user involved in comments of a given video should be regarded as *viewers* since they no longer constitute new environments for discussion. Building from existing literatures, we extend the subject of investigation beyond politics and societal agendas to product reviews on YouTube. Product reviews, as a video category, requires *content creators* to aggregate knowledge and personal experience to describe and evaluate products. Interestingly, this practice in product reviews also allows viewers to participate in discussions with minimal prior knowledge since the topic is already manifested in the content. This study sourced data from the YouTube Channel MKBHD, one of the most subscribed and viewed technology review agency that holds 16.3 million followers. This Channel is selected for its nuanced factual presentation and exceptional production quality. While other product review channels might generate greater viewership through shock, rhetoric or comedy, MKBHD contents propose well-reasoned recommendations supported by empirical evidence and interviews (Hurley 2022), thereby shifting viewers attention from the host's opinion to the product itself. Additionally, the Channel is able to produce review videos before products are available to the public, further inviting viewers to respond without owning a relevant device. An examination of the opinions formed under product reviews could reveal the audience's intention of media consumption, as well as whether viewers might reinforce opinions on a product by interacting with peers. We provided two guiding questions to highlight our research objectives:

**RQ1:** What type of comments are most engaged by viewers under selected product review videos?

By default, YouTube sorts comments by their popularity reflected through like/dislike counts and numbers of follow-up replies. Identifying and categorising popular comments provides us a glimpse of communication patterns within the community and a rationale for determining the similarity between comments, raising the question:

**RQ2:** To what extent, if *at all*, are sentiments of opinions from YouTube comments on selected product reviews *homogeneous*?

Here, we refer to the *homogeneity* of comments as a uniformity amongst reviewer sentiments regarding the product. The sentiment of viewers could be assigned with either *like, dislike*, or *neutral* to a product depending on their stance. Note that our study is not intended a market research on customer satisfaction; we examine comments from product reviews that arise regardless whether viewers possess them or not.

We initiate our research by crawling data from YouTube via YouTube Data Tools (YTDT) developed by Bernhard Rieder (2015), followed by combining quantitive content analysis (QCA) and sentiment analysis (SA) to interpret the communicational norms established by the community, and offer insights on viewer sentiments regarding their uniformity. These approaches have been extensively employed by media scholars, while we particularly build from QCA and SA methods proposed by Röchert et al. (2020) and Niederer (2016) to examine the comments.

**Method**

*Collecting and Cleaning Data*

Observing video contents produced by MKBHD on YouTube, we have selected 10 product review videos gaining the most comments. The selected videos features various distinguished technology companies such as Apple, Sony and Samsung, including a diversity of products ranging from smart phones to operating systems. A varied selection of products also warrants comment homogeneity solely consequential from viewer interactions by reducing bias introduced from repeated brand and product exposure. YTDT is then used to gather all comments generated from aforementioned contents, returning 10 datasets of comment texts and the amount of likes and replies received by each comment. The datasets are cleaned with automation to remove spams, short and meaningless phrases. Comments about the Channel and video production itself are also omitted since they do not relate to products.

*Classification and Labelling*

Remaining comments in datasets are then: 1) ranked by their like counts to analyse popular utterance formalities, 2) stratified into a dataset of 10 sub-sets containing 250 randomly selected comments from each video, totalling 2,500 entries across 10 videos, to evaluate comment homogeneity.

We initially collected 10 most liked comments from the selected product reviews (n=10) and created a dataset containing 100 entries to identify the components of their utterances. An observation of these entries reveals most comments poses opinions about products yet delivered with different argumentation. We hence created a coding scheme by the nature of argumentation in these comments:

**Fact**: the comment includes observable, verifiable information about the product without explicit expressions of favour nor disapproval.
*Example*: "2019: Air Pods Pro 250$; 2020: AirPods Max 550$; 2021: Air Pods Pro Max 1100$"
**Inference**: the comment formulates opinions from personal experience regarding products.

*Example*: "I'm currently using an S20 FE, and I have to admit that it s a great phone! I haven't experienced any lag at all, runs very smooth and I'm really enjoying it every single day."

**Judgement**: opinions proposed in the comment is solely supported by factual information.

*Example*: "the magsafe wireless charger just seems like wired charging with extra steps."

**Opinion**: comments expressing objective decisions with no further explanation.

*Example*: "Lol what? Making poor quality products is definitely on character for Apple."

With these categories in mind, we categorised the top comments across 10 videos. Table 1 features the distribution of argumentation types amongst the most favoured 100 comments:

**Table 1. Types of argumentation distributed in top comments**

|  | Fact | Inference | Judgement | Opinion | Total |
|---|---|---|---|---|---|
| **Occurance** | 5 | 67 | 8 | 20 | 100 |
| **Like counts** | 16543 | 473062 | 62351 | 399110 | 888715 |

The outcome from this classification suggests viewers particularly favours comments oriented from personal experiences. Furthermore, the numbers of likes received by each form of argumentation reveals a significant amount of users offered a nod to opinions of inference, that is to say, other viewers' personal experience.

*Sentiment Analysis*

This quantitative approach to popular comments, however, does not take into account the sentiment of viewer, meaning even though a large number of viewers affirms other users' experience, it is not possible to determine whether these experiences align with each other on their judgements. Here we introduce sentiment analysis (SA) to further examine the stance of viewers and the diversity of viewer opinions. With 2,500 comments randomly selected from 10 cleaned datasets, in addition to labelling their argumentation styles, sentiments of viewers are also annotated with three mutually exclusive stances: *positive*, *negative* and *others*. While positive and negative stances towards products are self-explanatory, comments that does not reflect any sentients are labelled *others*. This can include humour, insults and questions, seen in examples below:

*Example 1*: "im waiting for airpod max pro ultra studio hifi plus"

*Example 2*: "get rid of trash phone"

*Example 3*: "What games are updated for PS5 rn?"

Due to practical limitations of our research, the datasets were manually labelled on spreadsheets without computer-aided natural language processing. However, human annotation achieves accurate categorisations without the need to prepare training datasets for algorithms. The annotated results are summarised in Table 2.

**Table 2. Argumentation style and Sentiment distribution in selected video comments**

|  | Argumentation Style | | | | Sentiment | | |
|---|---|---|---|---|---|---|---|
| *Video ID* | Fact | Inference | Judgement | Opinion | Positive | Negative | Others |
| UdfSrJvqY_E | 4 | 32 | 110 | 104 | 0 | 221 | 29 |
| QtMzV73NAgk | 8 | 129 | 32 | 81 | 199 | 4 | 47 |
| azrdcp4yYas | 19 | 92 | 47 | 92 | 178 | 42 | 30 |
| Sx6dAx7dnXg | 0 | 70 | 72 | 108 | 19 | 184 | 47 |
| k1v7_zScivQ | 23 | 34 | 3 | 190 | 160 | 49 | 41 |
| ZLyDvABxGF0 | 0 | 132 | 63 | 55 | 197 | 26 | 27 |
| Gvvo6vUpJRc | 0 | 63 | 10 | 177 | 0 | 216 | 34 |
| ehv3zQAa9zM | 0 | 77 | 88 | 85 | 149 | 47 | 54 |
| X1b3C2081-Q | 1 | 133 | 24 | 92 | 159 | 26 | 65 |
| 8FpPSMIB4uA | 18 | 22 | 9 | 201 | 64 | 98 | 88 |

**Findings and Analysis**

*Norms amongst Review Comments*

The first stage of our research has demonstrated that while multiple expressions of opinions are present in product review comments, viewers tend to resort to *inference* and explicit *opinions* rather than *facts* and opinions deduced from them (*judgements*). In response to RQ1, our finding confirms the presence of dominant argumentative styles across multiple product review videos. The dominating expression can also be observed from Figure 1, whereas inference is notably more employed by viewers than other argumentation techniques.

**Figure 1. Distribution of Argumentation Styles among popular comments**

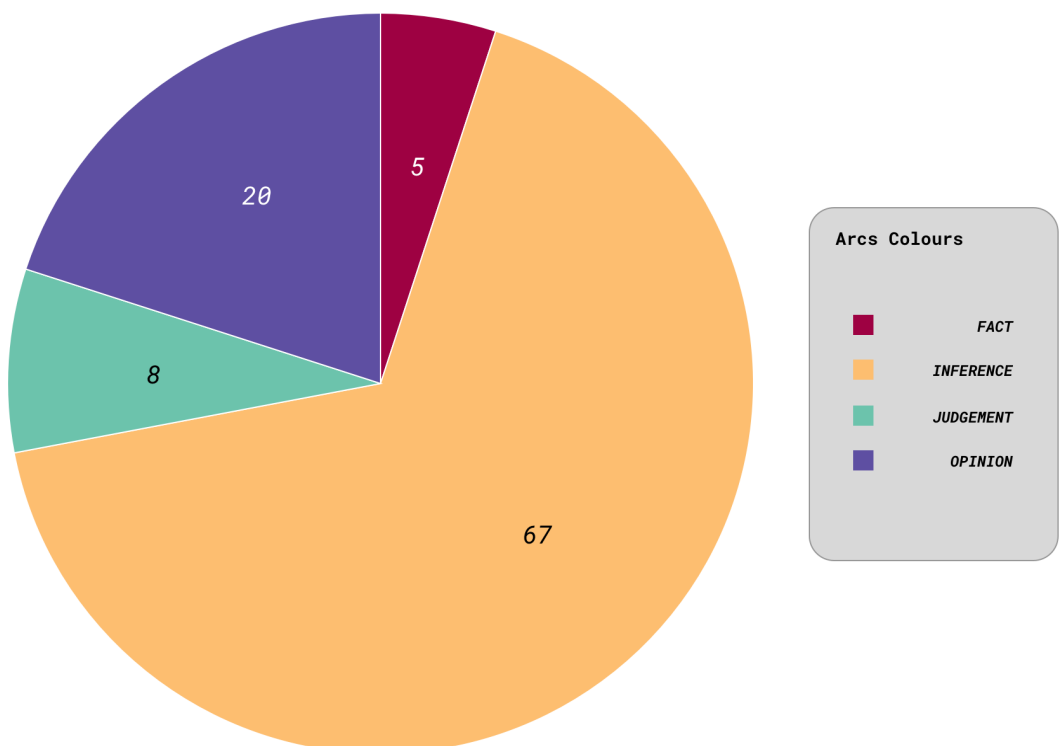

However, while opinions related with personal experience occurs the most frequently in our sample, insufficient evidence could be found to determine whether *inference* based comments are the most favoured by peer viewers. Table 3 present the average number of likes acquired from various argumentation styles. The parallel coordinate graph (**Figure 2**) produced based on Table 3 indicates no significant correlations between the style of comments and their gained popularity.

**Table 3. Popularity of different argumentation styles**

|  | Fact | Inference | Judgement | Opinion |
|---|---|---|---|---|
| **Like per occurrence** | 3308,6 | 7060,6 | 7793,9 | 19955,5 |

**Figure 2. Popularity of different argumentation styles in relation to occurrences**

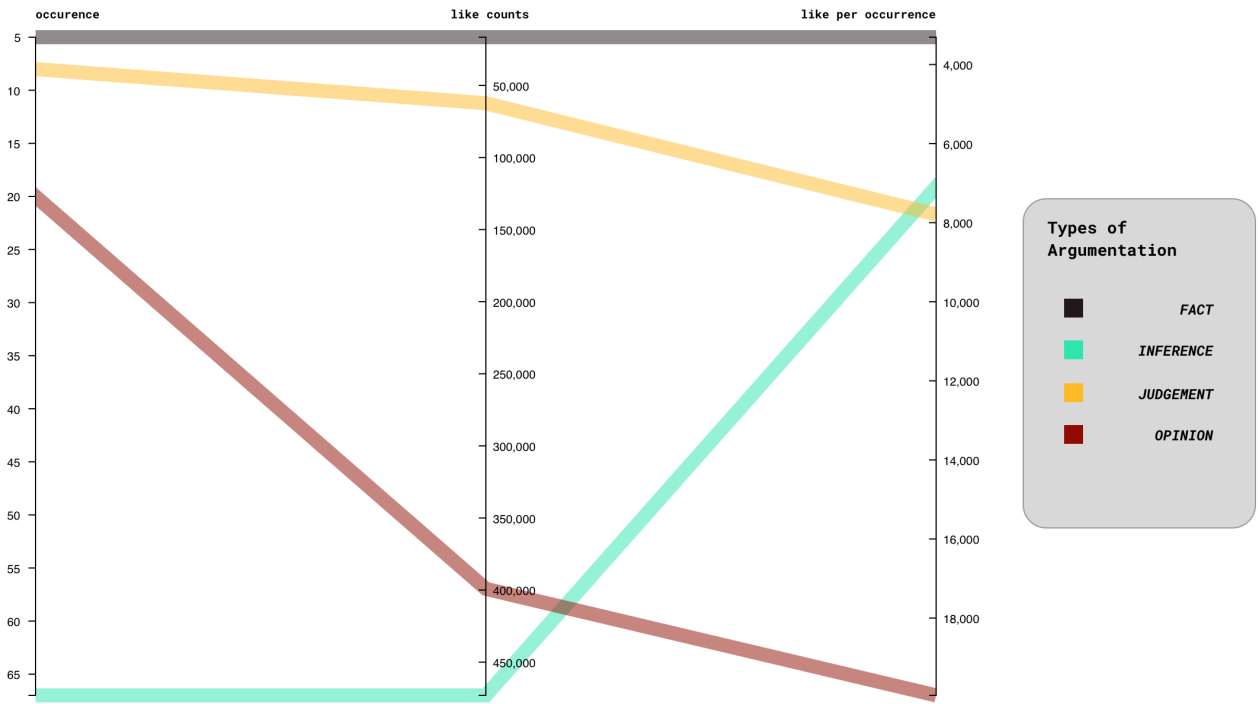

There are several factors that could contribute to the popularity of comments, such as the identity of the commenter, the use of rhetorical devices and when the comment was posted. YouTube algorithms might also prioritise certain comments which could foster popularity of certain comments. Considering the proprietary nature of YouTube, we were unable to determine the algorithmic influence on comment popularity.

*Visualising Comment Sentiment Homogeneity*

In order to answer RQ2 regarding comment homogeneity, we constructed Figure 3, a series of radar charts to illustrate the dispersion of sentients apropos of selected videos using labeled comments appeared in Table 2. The spokes of each plot represents possible sentiments: *positive* (top), *negative* (left) and *others* (right), whereas the coloured shapes represent the distribution of sentiments in a given video. The further they extend on a spoke, the more viewers express similar sentiments on a product discussed by the content creator. We discovered a significant sentiment inclination amongst selected viewer comments in 9 out of 10 videos sampled, indicating a strong likelihood of viewers repeatedly encountering similar stances to a product. Regarding the echo chamber phenomenon discussed earlier, it could be deduced that viewers of product reviews *do* frequently situate themselves in echo chambers that share and affirm similar opinions. Consequently, the popular notions in existing YouTube comments implicitly affect future viewers since they could reasonably predict whether their opinions would be acknowledged by other commentators, which further reinforces comment homogeneity for audiences old and new.

**Figure 3. Popularity of different argumentation styles in relation to occurrences**

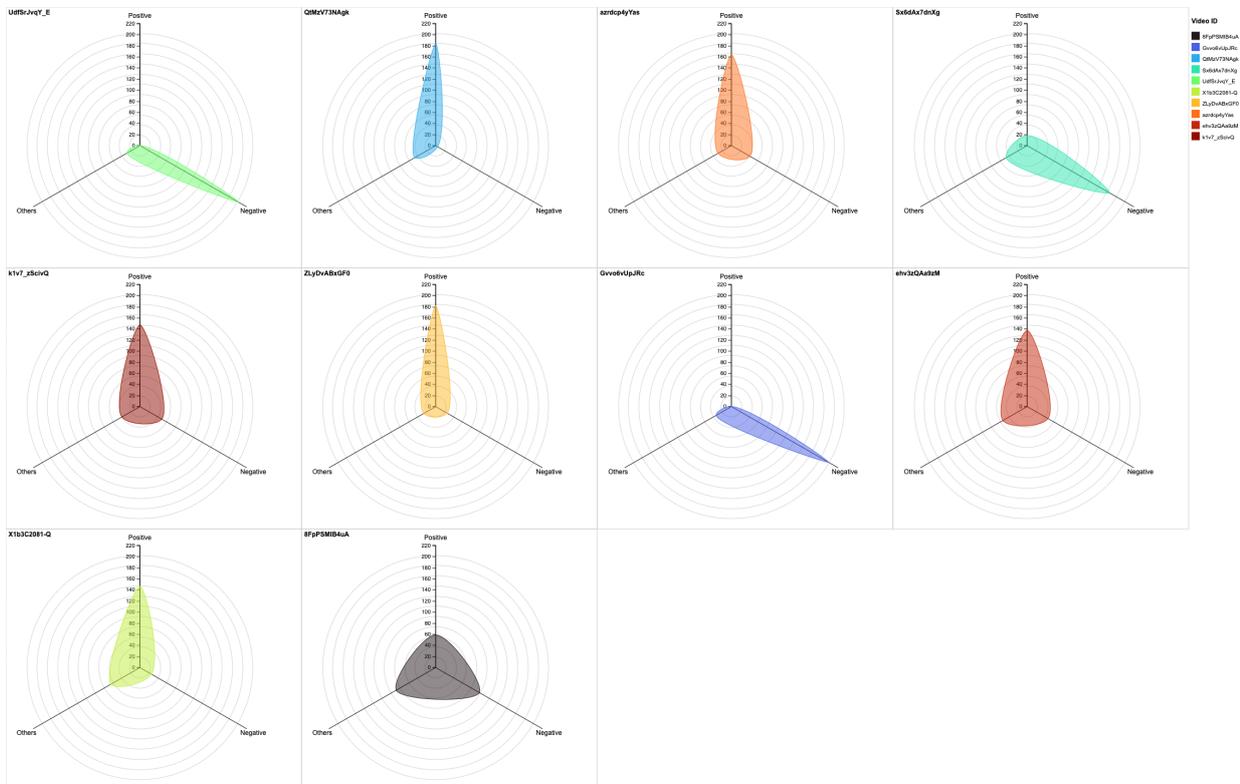

**Conclusion and Further Research**

By analysing YouTube comments through a combined approach of quantitative content analysis and sentiment analysis, we discovered that echo chambers could well exist outside traditional social media platforms and political agendas. Our investigation on product review comments proposes three major findings: 1) viewers of said contents implement several noticeable argumentation styles when conveying arguments; 2) while styles of argumentation used do not correlate to the popularity of comments, viewers are more likely to announce opinions and personal experiences; 3) product review comments exhibit highly identical sentiments that dwarf other stances, suggesting the likely presence of echo chambers in this content category which impacts not only current viewers but also the once to come.

However, our investigation of 2,500 comments in 10 product reviews from a Channel is far from a comprehensive view of how users and their comments affect each other on YouTube. Rather, this study serves as an overview to confirm the expansive presence of echo chambers across platforms and topics. Another limitation of our work arises from the error introduced by the time when comments were posted: earlier comments could benefit from information scarcity and receive more responses by peers regardless their message, resulting in comments not reflective of the viewers opinions on the products still becoming popular.

Future scholars could take advantage of customised applications to efficiently retrieve desired data from YouTube without using YTDT and minimise the labour during data cleaning, while natural language processing (NLP) and analytical tools such as SPSS would benefit an analysis on a larger dataset along with greater precision when evaluation results.

**Notes**

*Datasets and high-resolution graphs of this paper could be retrieved at:*

https://osf.io/j2wqm/?view_only=ff9eed6434da4ee69dd2464f1b520857